\documentclass[prb,11pt]{revtex4-1}

\usepackage[english]{babel}
\usepackage[intlimits]{amsmath}
\usepackage[dvips]{graphicx}
\usepackage{amsfonts}
\usepackage{amssymb}
\usepackage{graphicx}
\usepackage{bm}

\begin{document}

\begin{titlepage}

\title{{\bf Michaelis-Menten Relations for Complex Enzymatic Networks}}
\author{{\bf Anatoly B. Kolomeisky}}
\affiliation{Department of Chemistry, Rice University, Houston, TX 77005-1892, USA}

\begin{abstract}

All biological processes are controlled by complex systems of enzymatic chemical reactions. Although the majority of enzymatic networks have very elaborate structures, there are many experimental observations indicating that some turnover rates still follow a simple Michaelis-Menten relation with a hyperbolic dependence on a substrate concentration.  The original Michaelis-Menten mechanism has been derived as a steady-state  approximation for a single-pathway enzymatic chain. The validity of this mechanism for many complex enzymatic systems is surprising. To determine general conditions when this relation might be observed in experiments, enzymatic networks consisting of coupled parallel pathways are investigated theoretically. It is found that the Michaelis-Menten equation is satisfied for specific relations between chemical rates, and  it also corresponds to the situation with no fluxes between parallel pathways. Our results are illustrated for simple models. The importance of the Michaelis-Menten relationship and derived criteria for single-molecule experimental studies of enzymatic processes are discussed.

\end{abstract}

\maketitle

\end{titlepage}

\section{Introduction}

It is known that all chemical reactions in biological systems are catalyzed by protein enzymatic molecules.\cite{alberts,lodish} Fundamental understanding of cellular processes cannot be accomplished without determining  how involved enzymatic networks function. First catalytic mechanism that involves enzyme molecules has been proposed almost a century ago by Michaelis and Menten,\cite{MM} and since it became one of the most used and celebrated relations in biochemical and biophysical studies of natural phenomena. Similar relations between chemical rates and substrate concentrations have been observed in many enzymatic systems.\cite{segel,fersht,cook,english06} However, why the Michaelis-Menten (MM) mechanism, i.e., the hyperbolic dependence on the concentration of substrate, is working even for some very complex enzymatic networks is still not well understood.

The original MM mechanism has been derived as a steady-state approximation to a simple single-pathway enzymatic process with irreversible creation of product molecules as shown in a scheme at Fig. 1.\cite{houston} In addition, it was assumed that the initial concentration of enzyme was much smaller than the initial concentration of the substrate, and that not too many product molecules have been produced.\cite{houston} Then the rate of the  catalyzed reaction can be written in terms of the rate constants shown in Fig. 1 as
\begin{equation}
V=\frac{k_{2}c}{c+K_{M}},
\end{equation}
where $c$ is the concentration of substrate molecules, $K_{M}=(k_{2}+k_{-1})/k_{1}$ is known as the Michaelis constant and  the rate $k_{2}$ is called the catalytic rate. The rate of enzymatic reaction in the MM mechanism is proportional to the concentration of the substrate molecules at low concentrations while it saturates at high concentrations. Surprisingly, similar behavior is also observed experimentally in many enzymatic systems with much more complex topology of chemical transitions than one shown in Fig. 1.\cite{segel,fersht,cook,english06}  Several theoretical studies have been presented in order to understand why turnover rates in complex enzymatic networks might follow the simple MM relation.\cite{kou05,min06,gopich06,cao11} It has been argued that the MM equation will be satisfied for fluctuating enzymatic systems when conformational transition rates are very large or very low in comparison with catalytic rates, although with redefined definitions of catalytic rates and $K_{M}$ constants.\cite{kou05,min06}  A general approach to analyze enzymatic kinetics, based on the flux balance method, has been developed recently,\cite{cao11} and it suggests that the MM relation will not be observed when the detailed balance is broken. However, in all current approaches only enzymatic networks with several irreversible transitions have been considered. In addition, general expressions for the rate of product formation have not been derived, and explicit conditions on validity of the MM equation have not been obtained. 

In this paper, we present an alternative theoretical method of analyzing complex enzymatic networks based on  solving discrete-state stochastic master equations. This approach has been utilized successfully to describe  dynamic properties and mechanisms of motor proteins, and also in the analysis of other single-molecule experiments.\cite{KF00,kolomeisky00,kolomeisky01,stukalin06,AR,das09} Specifically, we  consider enzymatic networks consisting of coupled parallel pathways with all chemical transitions assumed to be reversible.  Our method allows us to describe explicitly conditions when the MM relation for the turnover rate is satisfied. These conditions are analogous to the requirement that the flux between parallel pathways vanishes at each state. Using simple models to explain our method we argue that all turnover rates behave qualitatively similar to the MM relation.

The paper is organized as follows. In Sec. II our theoretical method is presented for general coupled parallel enzymatic networks. Examples to illustrate obtained results are given in Sec. III. Finally, Sec. IV provides a summary of our findings.

\section{Theoretical Analysis}

As an example of complex enzymatic networks a system made of several coupled parallel biochemical pathways, as shown in Fig. 2, will be analyzed. We adopt here a single-molecule approach, and to simplify calculations all derivations will be made for the network of two coupled parallel pathways (see Fig. 2), although the analysis can be easily extended to many coupled parallel pathways. This situation corresponds to a single enzymatic molecule that can be found in one of two conformations, and it catalyzes the reaction of the substrate transformation to the corresponding product in both conformations but with different rates. 

It is assumed that in each pathway there are $N$ discrete chemical states per each enzymatic cycle. The enzyme molecule can be found in the state $i$  from which it can move forward or backward along the same reaction channel with rates $u_{i}$ or $w_{i}$, respectively, or it can change its conformation by moving to the state $i$ in the second pathway with the rate $\gamma_{i}$: see Fig. 2. Similarly, if the molecule is in the state $i$ of the second reaction channel it can make forward (backward) transition with the rate $\alpha_{i}$ ($\beta_{i}$), or it can change its conformational state by going to the first pathway with the rate $\delta_{i}$, as illustrated in Fig. 2. We also assume that the enzyme molecule binds to the substrate in the states $i=0$, i.e., the corresponding transition rates are proportional to the concentration of substrate molecules $c$, $u_{0}=k_{1}c$ and $\alpha_{0}=k_{2}c$. Let us define a function $P_{i}^{(k)}(t)$ as a probability to find the enzyme molecule in the state $i$ at the reaction channel $k$ ($k=1$ or 2) at time $t$. The temporal evolution of this enzymatic network can be described by a system of master equations,
\begin{equation}\label{me1}
\frac{d P_{i}^{(1)}(t)}{dt}=u_{i-1} P_{i-1}^{(1)}(t)+ w_{i+1} P_{i+1}^{(1)}(t)+ \delta_{i} P_{i}^{(2)}(t) -(u_{i}+w_{i}+\gamma_{i})P_{i}^{(1)}(t);
\end{equation}
\begin{equation}\label{me2}
\frac{d P_{i}^{(2)}(t)}{dt}=\alpha_{i-1} P_{i-1}^{(2)}(t)+ \beta_{i+1} P_{i+1}^{(2)}(t)+ \gamma_{i} P_{i}^{(1)}(t) -(\alpha_{i}+\beta_{i}+\delta_{i})P_{i}^{(2)}(t).
\end{equation}
Because of periodicity, it is convenient to define a new function,
\begin{equation}
R_{i}^{(k)}(t)=\sum_{l} P_{i+lN}^{(k)}(t),
\end{equation}
 with $l$ being any integer number. Then the conformational transition flux between the states $i$ ($i=0,1,\ldots, N-1$) in two pathways can be written as
\begin{equation}\label{flux}
J_{i}= \gamma_{i} R_{i}^{(1)}-\delta_{i} R_{i}^{(2)}.
\end{equation}

When the system reaches a stationary state ($t \rightarrow \infty$), Master equations (\ref{me1}) and (\ref{me2}) transform into the following expressions,
\begin{equation}\label{me_stat1}
0=u_{i-1} R_{i-1}^{(1)}+ w_{i+1} R_{i+1}^{(1)}(t) -J_{i} -(u_{i}+w_{i})R_{i}^{(1)};
\end{equation}
\begin{equation}\label{me_stat2}
0=\alpha_{i-1} R_{i-1}^{(2)}+ \beta_{i+1} R_{i+1}^{(2)} + J_{i} -(\alpha_{i}+\beta_{i})R_{i}^{(2)}.
\end{equation}
In addition, at large times total conformational flux between states in two pathways must disappear,
\begin{equation}\label{current}
\sum_{i=0}^{N-1} J_{i}=0.
\end{equation}
Now we can define new effective rates,
\begin{equation} \label{u_eff}
\widetilde{u_{i}}=u_{i}+\alpha_{i} \frac{\gamma_{i}}{\delta_{i}},
\end{equation}
and
\begin{equation} \label{w_eff}
\widetilde{w_{i}}=w_{i}+\beta_{i} \frac{\gamma_{i}}{\delta_{i}}.
\end{equation}
Using these definitions and the expression for the flux (\ref{flux}) one can obtain from Eq. (\ref{me_stat1})
\begin{equation}\label{eq_K}
0=\widetilde{u_{i-1}} R_{i-1}^{(1)}-\widetilde{w_{i}}R_{i}^{(1)}-K_{i-1}-\widetilde{u_{i}} R_{i}^{(1)}+\widetilde{w_{i+1}}R_{i+1}^{(1)} + K_{i},
\end{equation}
where a new function $K_{i}$ is defined as
\begin{equation}
K_{i}=\frac{\alpha_{i}}{\delta_{i}} J_{i}-\frac{\beta_{i+1}}{\delta_{i+1}} J_{i+1}.
\end{equation}
Since Eq. (\ref{eq_K}) must be satisfied for any chemical state $i$ it leads to
\begin{equation}\label{eq_K1}
\widetilde{u_{i}} R_{i}^{(1)}-\widetilde{w_{i+1}}R_{i+1}^{(1)} - K_{i}=C,
\end{equation}
where $C$ is some unknown constant. Using periodicity  Eq. (\ref{eq_K1}) can be easily solved, yielding
\begin{equation} \label{eq_Ri}
R_{i}^{(1)}=\frac{C r_{i}+r_{i}(K)}{1-\Gamma},
\end{equation}
with
\begin{equation}
\Gamma=\prod_{i=0}^{N-1} \frac{\widetilde{w_{i}}}{\widetilde{u_{i}}};
\end{equation}
\begin{equation}
r_{i}=\frac{1}{\widetilde{u_{i}}}\left(1+ \sum_{k=1}^{N-1} \prod_{j=1}^{k} \frac{\widetilde{w_{j+1}}}{\widetilde{u_{j+1}}}\right);
\end{equation}
and
\begin{equation}
r_{i}(K)=\frac{1}{\widetilde{u_{i}}}\left(K_{i}+ \sum_{k=1}^{N-1} K_{i+k}\prod_{j=1}^{k} \frac{\widetilde{w_{j+1}}}{\widetilde{u_{j+1}}}\right).
\end{equation}
The unknown constant $C$ can be found from the normalization condition, namely,
\begin{equation}
\sum_{i=0}^{N-1} \left(R_{i}^{(1)}+R_{i}^{(2)}\right)=1,
\end{equation}
which produces
\begin{equation}\label{const}
C=\frac{(1+L)(1-\Gamma)-S(K)}{S},
\end{equation}
where new auxiliary functions are defined as
\begin{equation}
L=\sum_{i=0}^{N-1} \frac{J_{i}}{\delta_{i}};
\end{equation}
\begin{equation}
S=\sum_{i=0}^{N-1}\left(1+\frac{\gamma_{i}}{\delta_{i}}\right) r_{i},
\end{equation}
and
\begin{equation}
S(K)=\sum_{i=0}^{N-1}\left(1+\frac{\gamma_{i}}{\delta_{i}}\right) r_{i}(K).
\end{equation}
Substituting  Eq. (\ref{const}) into Eq. (\ref{eq_Ri}) it can be shown that
\begin{equation}
R_{i}^{(1)}= \left( \frac{1+L}{R}-\frac{R(K)}{R(1-\Gamma)} \right)r_{i}+\frac{r_{i}(K)}{(1-\Gamma)}.
\end{equation}
These expression for the probability to be found in the state $i$ of the first reaction channel at steady-state conditions allow us to compute all dynamic properties of the system. The turnover rate for this enzymatic network,
\begin{equation}
V=\sum_{i=0}^{N-1} (u_{i}-w_{i}) R_{i}^{(1)}+\sum_{i=0}^{N-1} (\alpha_{i}-\beta_{i}) R_{i}^{(2)},
\end{equation}
which can be written as
\begin{equation}
V=\sum_{i=0}^{N-1} (\widetilde{u_{i}}-\widetilde{w_{i}}) R_{i}^{(1)}-\sum_{i=0}^{N-1} K_{i}.
\end{equation}
Then it can be shown that
\begin{equation}
\sum_{i=0}^{N-1} (\widetilde{u_{i}}-\widetilde{w_{i}}) r_{i} = N(1-\Gamma),
\end{equation}
and
\begin{equation}
\sum_{i=0}^{N-1} (\widetilde{u_{i}}-\widetilde{w_{i}}) r_{i}(K) = (1-\Gamma)\sum_{i=0}^{N-1}K_{i}.
\end{equation}
It leads to a compact expression for the chemical reaction rate,
\begin{equation}
V=\frac{N(1-\Gamma)}{S}+\frac{N(1-\Gamma)L}{S}+\frac{NR(K)}{S}.
\end{equation}

We can repeat derivations if we choose another set of effective rates,
\begin{equation}
\widetilde{\alpha_{i}}=\alpha_{i}+u_{i} \frac{\delta_{i}}{\gamma_{i}},
\end{equation}
and
\begin{equation}
\widetilde{\beta_{i}}=\beta_{i}+w_{i} \frac{\delta_{i}}{\gamma_{i}}.
\end{equation}
Comparing them with Eqs. (\ref{u_eff}) and (\ref{w_eff}) we conclude that
\begin{equation}
\frac{\widetilde{u_{i}}}{\widetilde{\alpha_{i}}}=\frac{\widetilde{w_{i}}}{\widetilde{\beta_{i}}}=\frac{\gamma_{i}}{\delta_{i}}.
\end{equation}
Again the equation for the turnover rate can be derived with this set of effective rates. Combining both expression, we arrive to the final formula for the reaction speed of the enzymatic process presented in Fig. 2,
\begin{equation}\label{eq_rate}
V=\frac{N(1-\Gamma)}{S}+\frac{N(1-\Gamma)}{2S}\sum_{i=0}^{N-1}J_{i}\left(\frac{1}{\delta_{i}}-\frac{1}{\gamma_{i}}\right)+\frac{NS(M)}{2S},
\end{equation}
with
\begin{equation}
r_{i}(M)=\frac{1}{\widetilde{u_{i}}}\left(M_{i}+ \sum_{k=1}^{N-1}M_{i+k}\prod_{j=1}^{k} \frac{\widetilde{w_{j+1}}}{\widetilde{u_{j+1}}}\right),\end{equation}
\begin{equation}
S(M)=\sum_{i=0}^{N-1}\left(1+\frac{\gamma_{i}}{\delta_{i}}\right) r_{i}(M),
\end{equation}
and  functions $M_{i}$ are defined as
\begin{equation}
M_{i}=\left(\frac{\alpha_{i}}{\delta_{i}}-\frac{u_{i}}{\gamma_{i}}\right) J_{i}-\left(\frac{\beta_{i+1}}{\delta_{i+1}} - \frac{w_{i+1}}{\gamma_{i+1}}\right) J_{i+1}.
\end{equation}

Eq. (\ref{eq_rate}) is an exact expression for the turnover rate for general enzymatic network made of coupled parallel pathways.  It depends on chemical rates for individual transitions  as well as conformational fluxes $J_{i}$. It should be noted that these fluxes are not independent parameters. Explicit forms of conformational fluxes can be found utilizing Eqs. (\ref{flux}) and also from the balance of total fluxes through each chemical state,\cite{stukalin06}
\begin{equation}
u_{i-1} R_{i-1}^{(1)}-w_{i} R_{i}^{(1)}= J_{i}+ u_{i} R_{i}^{(1)}-w_{i+1} R_{i+1}^{(1)},
\end{equation}
\begin{equation}
\alpha_{i-1} R_{i-1}^{(2)}-\beta_{i} R_{i}^{(2)}+ J_{i}= \alpha_{i} R_{i}^{(2)}-\beta_{i+1} R_{i+1}^{(2)}.
\end{equation}

It is convenient to analyze the expression (\ref{eq_rate})  for the overall chemical rate of the coupled enzymatic network if it follows or not the MM dependence. It can be easily shown that the first term has a hyperbolic dependence on the concentration of the substrate and it does not depend on conformational fluxes. At the same time, the second and the third terms  do depend on $J_{i}$ and they are  not satisfying the MM relation for any non-zero conformational fluxes. Thus the condition for the turnover rate to follow the MM dependence is when last two terms in Eq. (\ref{eq_rate}) vanish, which can only take place when conformational fluxes disappear. This condition can be written explicitly in terms of  relations between chemical rates,
\begin{equation}\label{relations}
\frac{u_{i}}{\alpha_{i}}=\frac{w_{i}}{\beta_{i}}=\frac{\gamma_{i}}{\delta_{i}},
\end{equation} 
or in the different form as
\begin{equation}\label{relations1}
\frac{u_{i}}{\gamma_{i}}=\frac{\alpha_{i}}{\delta_{i}}, \quad \frac{w_{i}}{\gamma_{i}}=\frac{\beta_{i}}{\delta_{i}}.
\end{equation} 
This is the main result of our work. A general coupled parallel enzymatic network will follow the MM relation when rates of coupled pathways satisfy the equation (\ref{relations}) for each chemical state. This is a clear and explicit criterion on the MM dependence of the enzymatic cycle in such complex systems. 

All previously discussed situations of the MM behavior for the turnover rate in enzymatic networks\cite{kou05,min06,cao11} can be described by our general approach. It has been argued that conformationally fluctuating enzymes still obey the MM relationship for 1) quasi-static conditions (when conformational rates are very slow in comparison with other chemical transitions), and for 2) quasi-equilibrium conditions (when conformational rates are very fast). The last case has been also analyzed in Ref. \cite{stukalin06}. The first situation corresponds to $\gamma_{i} \ll 1$ and $\delta_{i} \ll 1$. Then Eqs. (\ref{relations1}) are  valid for any value of other chemical  rates (as long as they are larger than conformational transitions rates).  The second case describes very large $\gamma_{i}$ and $\delta_{i}$, and again Eqs. (\ref{relations1}) predict that  it might happen for any value of chemical rates $u_{i}$, $w_{i}$, $\alpha_{i}$ and $\beta_{i}$. It was also found\cite{min06} that the MM relation still holds when only one group of the conformationally-related states $i$ is found in quasi-static or quasi-equilibrium conditions. Since $N=2$ periodic system has been used for analyzing fluctuating enzymes, then from Eq. (\ref{current}) one can easily conclude that there are two conformational fluxes, $J_{1}$ and $J_{2}$, and they are related via $J_{1}+J_{2}=0$. If one of them goes to zero the second flux must also vanish, and the MM relationship is recovered. Cao has argued that the MM dependence is observed when the detailed balance is not broken.\cite{cao11} For the system of coupled parallel pathways it means that the overall circular current in each loop (see Fig. 2) is zero. The ratio of clockwise and counterclockwise currents for any loop can be written with the help of Eqs. (\ref{relations}) as
\begin{equation} 
\frac{u_{i} \gamma_{i+1} \beta_{i+1} \delta_{i}}{\alpha_{i} \delta_{i+1} w_{i+1} \gamma_{i}}=\frac{\gamma_{i+1} \beta_{i+1}}{\delta_{i+1} w_{i+1}}=1,
\end{equation}
which indicates that the loop current vanishes when the MM relation is observed.

It is important to note also that criteria given in Eqs. (\ref{relations}) suggest that the MM dependence is found for systems where each state has similar free-energy landscape near each chemical state. Then the enzymatic network of coupled parallel reaction channels can be effectively viewed as a single enzymatic pathway with properly rescaled transitions rates which by definition follow the MM behavior.

\section{Examples}

To illustrate our theoretical method let us present some explicit results for the $N=2$ system which is the most relevant for analyzing fluctuating enzyme systems.\cite{MM,segel,cook,kou05,min06}  The dynamic properties of enzyme molecule in such system has been already obtained.\cite{stukalin06} The general formula for the turnover rate is given by\cite{stukalin06}
\begin{eqnarray}\label{eqV2}
V&=&  (1/\Omega) \{  k_{1}c u_{1}-w_{0}w_{1})[(k_{2}c+\beta_{0})\delta_{1}+(\alpha_{1}+\beta_{1})\delta_{0}+\delta_{0}\delta_{1}] \\ \nonumber
& & + (k_{2}c \alpha_{1}-\beta_{0}\beta_{1})[(k_{1}c+w_{0})\gamma_{1}+(u_{1}+w_{1})\gamma_{0}+\gamma_{0}\gamma_{1}]\\ \nonumber
& &  +(k_{1}c\alpha_{1}-w_{0}\beta_{1})\delta_{0}\gamma_{1}+ (k_{2}c u_{1}-\beta_{0} w_{1})\gamma_{0}\delta_{1} \},
\end{eqnarray}  
where the function $\Omega$ is defined as
\begin{eqnarray}\label{Omega}
\Omega & = & (k_{1}c +w_{0}+\gamma_{0})[(k_{2}c+\beta_{0})\delta_{1}+(\alpha_{1}+\beta_{1})\gamma_{1}] \\ \nonumber
& & + (u_{1} +w_{1}+\gamma_{1})[(k_{2}c+\beta_{0})\gamma_{0}+(\alpha_{1}+\beta_{1})\delta_{0}] \\ \nonumber
& & + (k_{2}c +\beta_{0}+\delta_{0})[(k_{1}c+w_{0})\gamma_{1}+ (u_{1}+w_{1})\delta_{1}] \\ \nonumber
& & + (\alpha_{1} +\beta_{1}+\delta_{1})[(k_{1}c+w_{0})\delta_{0}+ (u_{1}+w_{1})\gamma_{0}].
\end{eqnarray}
The conformational flux is equal to
\begin{equation}\label{J2}
J_{0}=-J_{1}=J= (1/\Omega) [(u_{1}+w_{1})(k_{2}c+\beta_{0})\gamma_{0}\delta_{1}-(k_{1}c+w_{0})(\alpha_{1}+\beta_{1})\gamma_{1}\delta_{0}].
\end{equation}
When the conformational fluxes are zero the turnover rate has the MM dependence,
\begin{equation}
V=\frac{(k_{1}\delta_{0}+k_{2}\gamma_{0})(u_{1}\delta_{1}+\alpha_{1}\gamma_{1})c-(w_{0}\delta_{0}+\beta_{0}\gamma_{0})(w_{1}\delta_{1}+\beta_{1}\gamma_{1})}{[(k_{1}c+w_{0})\delta_{0}+(k_{2}c+\beta_{0})\gamma_{0}](\gamma_{1}+\delta_{1})+ [u_{1}+w_{1})\delta_{1}+(\alpha_{1}+\beta_{1})\gamma_{1}](\gamma_{0}+\delta_{0})}.
\end{equation}

Reaction rates for different sets of parameters are presented in Fig. 3. The dependence of all enzymatic rates on the substrate concentration is qualitatively similar: increasing function at small $c$ and saturation to a constant value for $c \gg 1$. There is nothing special in the MM relation (black curve in Fig. 3). This can be easily understood because the turnover rate is equal to the ratio of polynomial functions of $c$. For the MM case both numerator and denominator are just linear functions. At large concentrations of the substrate molecules transitions from the states  $i=1$ become rate-limiting leading to effectively concentration-independent behavior. One could also observe that changing transition rates for $i=0$ does not affect much overall enzymatic rates in comparison with the MM relation (green and blue curves in Fig. 3). However, modifying transition rates associated with $i=1$ can significantly alter reaction rates (red and brown curves in Fig. 3). In addition, analyzing all reaction rates curves we  conclude that many experimental observations for systems that could analyzed by $N=2$ coupled periodic models might be easily assumed to follow effectively the MM relation, especially if experimental errors are taken into account.  

The conformational fluxes for different sets of parameters are shown in Fig. 4. In the MM case the $J=0$ as expected, while in other cases the complex behavior is observed. At large concentrations all conformational fluxes tend to zero because in this case transitions from $i=1$ are becoming rate limiting, and they do depend on $c$. In some situations the magnitude of the conformational flux reaches a maximum for the specific concentration of the substrate molecules.

\section{Summary and Conclusions}

A new theoretical method of analyzing complex enzymatic networks is developed. It is based on solving explicitly master equations for discrete-state stochastic models. Applying this approach for systems made of coupled parallel enzymatic pathways, we derived explicit  criteria on when the MM behavior might be observed. It leads to special relations between all chemical rates. This method allows to explain all previous theoretical observations on the validity of the MM relationship in complex enzymatic systems. Theoretical analysis is illustrated for simple models relevant for conformationally fluctuating enzyme molecules. It is found that general behavior of enzymatic rates is qualitatively similar to the MM equation. Our theoretical method argues that turnover rates can be viewed as a ratio of polynomial functions of the substrate concentrations. The MM relationship is observed when these polynomials are linear functions of concentrations.        

The presented theoretical method might be a powerful tool for analyzing single-molecule experiments since it allows to compute all dynamic properties for complex enzymatic networks. It will be interesting to test experimentally developed criteria for coupled enzymatic pathways. It will be also important to extend this approach to other enzymatic systems and for computations of other experimentally observed quantities such as dwell-time distributions and diffusion constants.

\section*{Acknowledgments}

The author would like to acknowledge the support from the Welch Foundation (Grant  C-1559), the  U.S. National Science Foundation (Grant  ECCS-0708765), and from U.S. National Institute of Health (Grant R01GM094489). The author also acknowledges useful discussions with Jianshu Cao.

\newpage

\noindent {\bf Figure Captions:} \\
\vspace{5mm}

\noindent Fig. 1. A reaction scheme for the simple enzymatic process that can be described by the Michael-Menten (MM) mechanism. An enzyme molecule $E$ reacts reversibly with a substrate molecule $S$ to produce an intermediate molecule $X$ which can irreversibly transition to a product molecule $P$ and the enzyme $E$ with corresponding rates.   

\vspace{5mm}

\noindent Fig. 2. Enzymatic network with two coupled parallel pathways. Both pathways have $N$ discrete states per each enzymatic cycle. The enzyme molecule in the state $i$ can transition forward (backward) with the rate $u_{i}$ ($w_{i}$) if found in the pathway 1, while in the second channel the forward (backward) rates are $\alpha_{i}$ ($\beta_{i}$) (with $i=0,1,\ldots, N-1$). The conformational transitions between two $i$-th states in different pathways are given by rates $\gamma_{i}$ and $\delta_{i}$.  

\vspace{5mm}

\noindent Fig. 3. The turnover reaction rate as a function of the concentration of substrate molecules for $N=2$ periodic coupled parallel enzymatic network. Calculations have been performed utilizing Eq. (\ref{eqV2}) for the following parameters for all curves: $k_{1}=10$ $\mu$M$^{-1}$ s$^{-1}$, $k_{2}=1$ $\mu$M$^{-1}$ s$^{-1}$, $u_{1}=5$ s$^{-1}$, $w_{1}=\gamma_{0}=1$ s$^{-1}$, $\beta_{0}=\beta_{1}=0.1$ s$^{-1}$ and $\delta_{1}=0.5$  s$^{-1}$. In addition, for the brown curve we used $\alpha_{1}=5$ s$^{-1}$, $\delta_{0}=0.1$ s$^{-1}$, $\gamma_{1}=5$ s$^{-1}$ and $w_{0}=1$ s$^{-1}$; for the green curve we used $\alpha_{1}=0.5$ s$^{-1}$, $\delta_{0}=1$ s$^{-1}$, $\gamma_{1}=5$ s$^{-1}$ and $w_{0}=1$ s$^{-1}$; for the black curve we used $\alpha_{1}=0.5$ s$^{-1}$, $\delta_{0}=0.1$ s$^{-1}$, $\gamma_{1}=5$ s$^{-1}$ and $w_{0}=1$ s$^{-1}$; for the blue curve we used $\alpha_{1}=0.5$ s$^{-1}$, $\delta_{0}=0.1$ s$^{-1}$, $\gamma_{1}=5$ s$^{-1}$ and $w_{0}=10$ s$^{-1}$; and  for the red curve we used $\alpha_{1}=0.5$ s$^{-1}$, $\delta_{0}=0.1$ s$^{-1}$, $\gamma_{1}=50$ s$^{-1}$ and $w_{0}=1$ s$^{-1}$. The Michaelis-Menten case is described by the black curve.

\vspace{5mm}

\noindent Fig. 4. The conformational flux as a function of the concentration of substrate molecules for $N=2$ periodic coupled parallel enzymatic network. Calculations have been performed utilizing Eq. (\ref{J2}) for the following parameters for all curves: $k_{1}=10$ $\mu$M$^{-1}$ s$^{-1}$, $k_{2}=1$ $\mu$M$^{-1}$ s$^{-1}$, $u_{1}=5$ s$^{-1}$, $w_{1}=\gamma_{0}=1$ s$^{-1}$, $\beta_{0}=\beta_{1}=0.1$ s$^{-1}$ and $\delta_{1}=0.5$  s$^{-1}$. In addition, for the brown curve we used $\alpha_{1}=5$ s$^{-1}$, $\delta_{0}=0.1$ s$^{-1}$, $\gamma_{1}=5$ s$^{-1}$ and $w_{0}=1$ s$^{-1}$; for the green curve we used $\alpha_{1}=0.5$ s$^{-1}$, $\delta_{0}=1$ s$^{-1}$, $\gamma_{1}=5$ s$^{-1}$ and $w_{0}=1$ s$^{-1}$; for the blue curve we used $\alpha_{1}=0.5$ s$^{-1}$, $\delta_{0}=0.1$ s$^{-1}$, $\gamma_{1}=5$ s$^{-1}$ and $w_{0}=10$ s$^{-1}$; and  for the red curve we used $\alpha_{1}=0.5$ s$^{-1}$, $\delta_{0}=0.1$ s$^{-1}$, $\gamma_{1}=50$ s$^{-1}$ and $w_{0}=1$ s$^{-1}$. For parameters describing the MM relationship the conformational flux is equal to zero.

\newpage

\begin{figure}[ht]
\begin{center}
\unitlength 1in
\begin{picture}(3.0,4.0)
  \resizebox{3.375in}{1.4in}{\includegraphics{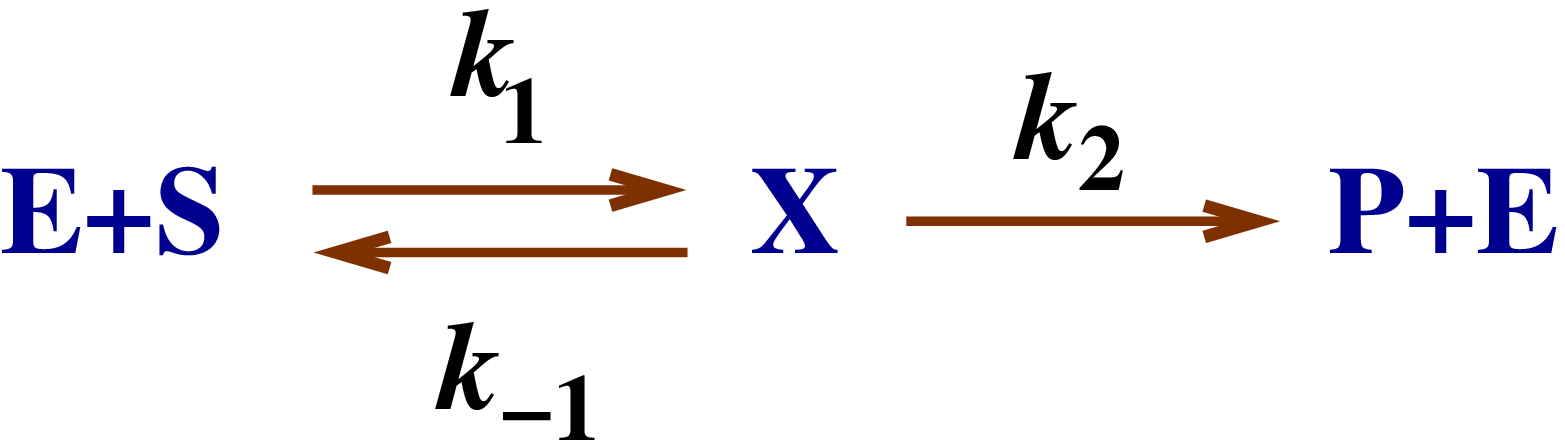}}
\end{picture}
\vskip 1in
 \begin{Large} Figure 1. Kolomeisky \end{Large}
\end{center}
\end{figure}

\newpage

\begin{figure}[ht]
\begin{center}
\unitlength 1in
\begin{picture}(3.0,4.0)
  \resizebox{3.375in}{1.7in}{\includegraphics{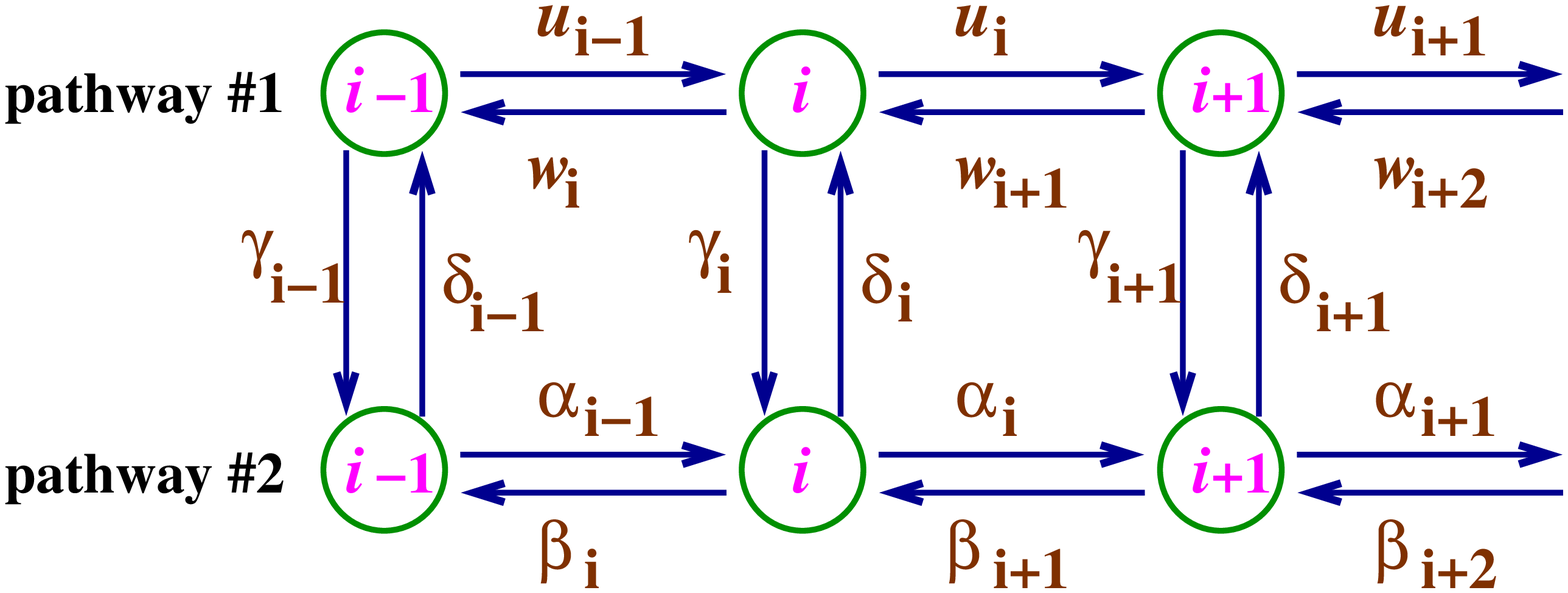}}
\end{picture}
\vskip 1in
 \begin{Large} Figure 2. Kolomeisky \end{Large}
\end{center}
\end{figure}

\newpage

\begin{figure}[ht]
\begin{center}
\unitlength 1in
\begin{picture}(3.0,4.0)
  \resizebox{3.375in}{3.375in}{\includegraphics{Fig3.eps}}
\end{picture}
\vskip 1in
 \begin{Large} Figure 3. Kolomeisky \end{Large}
\end{center}
\end{figure}

\newpage

\begin{figure}[ht]
\begin{center}
\unitlength 1in
\begin{picture}(3.0,4.0)
  \resizebox{3.375in}{3.375in}{\includegraphics{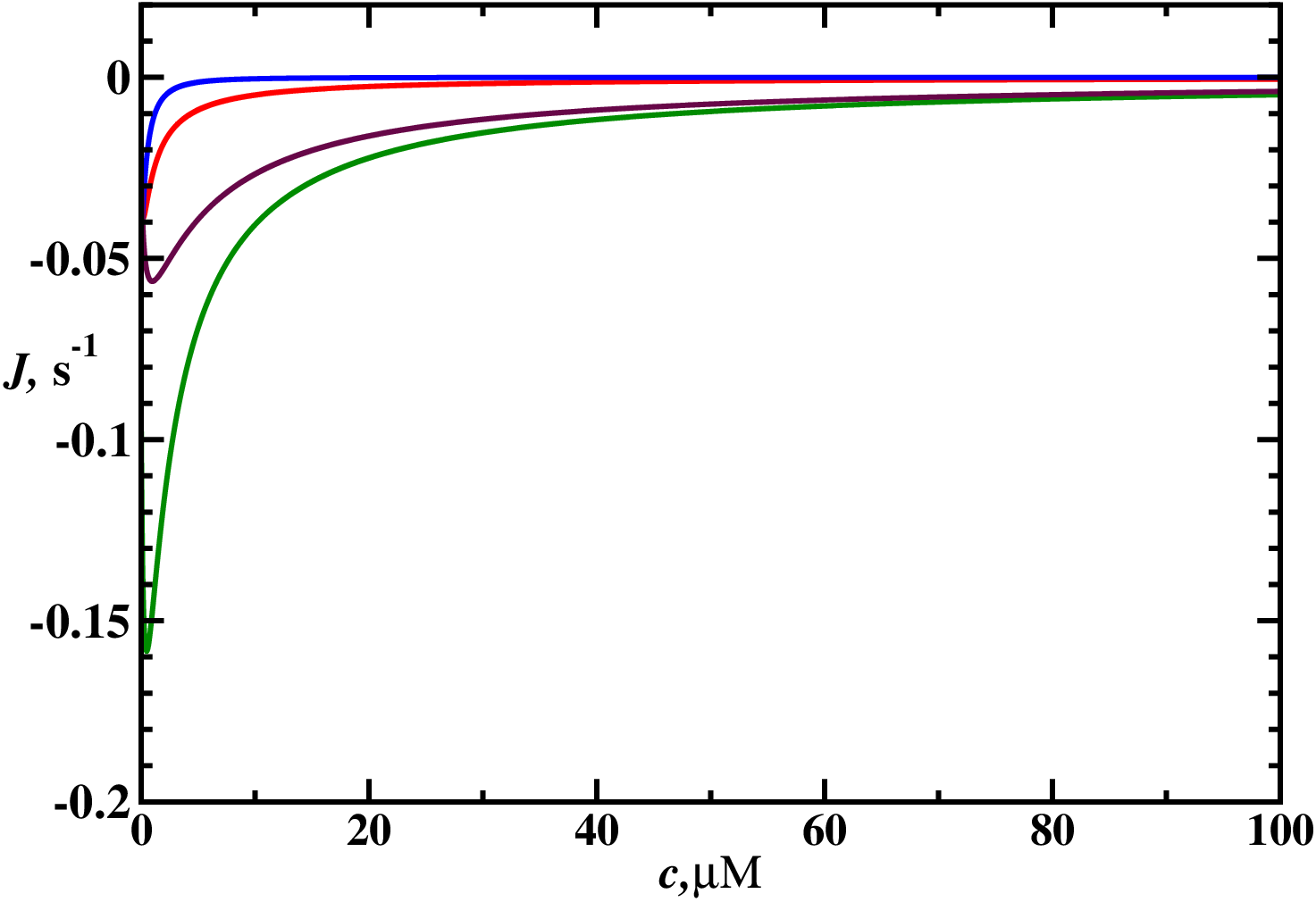}}
\end{picture}
\vskip 1in
 \begin{Large} Figure 4. Kolomeisky \end{Large}
\end{center}
\end{figure}

\end{document}